\documentclass[12pt]{article}

\usepackage{graphicx}
\usepackage{amsmath}
\usepackage{amssymb}
\allowdisplaybreaks

\begin{document}

\baselineskip=17.5pt plus 0.2pt minus 0.1pt

\renewcommand{\theequation}{\arabic{equation}}
\renewcommand{\thefootnote}{\fnsymbol{footnote}}
\makeatletter
\def\CR{\nonumber \\}
\def\pt{\partial}
\def\be{\begin{equation}}
\def\ee{\end{equation}}
\def\bea{\begin{eqnarray}}
\def\eea{\end{eqnarray}}
\def\eq#1{(\ref{#1})}
\def\la{\langle}
\def\ra{\rangle}
\def\hyp{\hbox{-}}


\begin{titlepage}
\title{\hfill\parbox{4cm}{ \normalsize YITP-07-32}\\
\vspace{1cm} The fluctuation spectra around a Gaussian classical solution of a tensor model and the general relativity}
\author{
Naoki {\sc Sasakura}\thanks{\tt sasakura@yukawa.kyoto-u.ac.jp}
\\[15pt]
{\it Yukawa Institute for Theoretical Physics, Kyoto University,}\\
{\it Kyoto 606-8502, Japan}}
\date{}
\maketitle
\thispagestyle{empty}

\begin{abstract}
\normalsize
Tensor models can be interpreted as theory of dynamical fuzzy spaces.
In this paper, I study numerically the fluctuation spectra around a
Gaussian classical solution of a tensor model, which represents a fuzzy flat space in arbitrary dimensions. 
It is found that the momentum distribution 
of the low-lying low-momentum spectra is 
in agreement with that of the metric tensor modulo 
the general coordinate transformation 
in the general relativity at least 
in the dimensions studied numerically, i.e. one to four dimensions.  
This result suggests that the effective field theory around the solution is 
described in a similar manner as the general relativity.   
\end{abstract}
\end{titlepage}

\section{Introduction}
\label{sec:intro}
There have been various discussions concerning the fuzziness of 
spacetime \cite{Garay}-\cite{Maziashvili:2006ey}.
In spite of the absence of fundamental theory of spacetime, 
it is generally argued that quantum gravitational fluctuations 
invalidate the reality of the classical notion of spacetime, i.e. 
a smooth continuous manifold in the general relativity.  
Therefore, it is widely believed that this classical notion should  
be replaced in some way with a new quantum notion.
Another unsatisfactory feature of this classical notion is that 
a spacetime and the gravitational degrees of freedom, i.e. a metric tensor, are 
independent distinct objects, i.e. in principle, it is allowed to 
consider a spacetime without a metric tensor.
This kind of insufficiency seems to exist also at least in the present formulation of string theory, i.e.
a string and a spacetime are quite different physical objects, although they are tightly related in several ways, 
such as the restrictions of spacetime dimensions. 
Therefore it is desirable that a new notion of spacetime 
contains a spacetime itself and the physical degrees of freedom 
in an inseparable fashion.  
The construction of a quantum theory of spacetime or quantum gravity would
essentially be equivalent
to finding such a new notion of spacetime which contains inseparably 
the general relativity in its classical limit.  

An interesting candidate for such a new notion of spacetime comes from 
fuzzy spaces. The central idea is the equivalence between a spacetime and 
the algebra of functions
on it. 
For a usual space, this is the implication of Gelfand-Naimark theorem \cite{connes}. 
Considering a deformation of an algebra of functions on a usual spacetime, 
a fuzzy spacetime can be obtained. 
When the deformed algebra is noncommutative, it defines a noncommutative 
spacetime \cite{connes}-\cite{Doplicher:1994tu}.
The deformed algebra can even be nonassociative 
\cite{Jackiw:1984rd}-\cite{Ho:2007vk}. This direction of deformation to
nonassociativity would be important in allowing an algebra to contain the information of 
geometric structures of a fuzzy space\footnote{This was partially insisted 
in the previous work \cite{Sasakura:2006pq} by the present author. 
In noncommutative spacetime,
the geometric properties are contained in the Laplacian rather than in the algebra 
\cite{connes}.}.

In view of the above idea of fuzzy spaces, 
a dynamical theory of spacetime may be obtained 
by a dynamical theory of fuzzy spaces. 
A fuzzy space is defined by an algebra, 
and an algebra can 
be given by a rank-three tensor ${C_{ab}}^c$, 
which defines the multiplication rule 
among the algebraic elements $f_a$ as $f_a f_b={C_{ab}}^c f_c$. 
Therefore it would be possible that a dynamical theory of 
a rank-three tensor ${C_{ab}}^c$, i.e. a tensor model,
can be regarded as a dynamical theory of spacetime.
This direction was pursued by the present author
\cite{Sasakura:2005js}-\cite{Sasakura:2006pq}, and a number of physically 
interesting classical solutions of tensor models were found numerically, 
which represent commutative nonassociative fuzzy spheres and tori in various dimensions
\cite{Sasakura:2006pq}.

It is interesting to note that tensor models have also appeared 
in some other contexts of quantum gravity.
Tensor models were considered as analytic expressions describing the dynamical 
triangulation model of spacetime in more than two dimensions and 
topological lattice gravities
\cite{Ambjorn:1990ge}-\cite{Ooguri:1992eb}. 
They also appeared in relation with the loop quantum gravity 
\cite{DePietri:1999bx}-\cite{Freidel:2005cg}.
The main difference of these previous approaches from the present one is the 
interpretation of tensor models. In the previous approaches, the Feynman graphs 
of the amplitudes of tensor models correspond one-to-one 
to the dual diagrams of triangulated manifolds. Computations of tensor models 
under this interpretation need 
essentially quantum treatment of tensor models, which, however, requires
future technical developments. On the other hand, under the new interpretation, 
a classical solution corresponds to a 
background spacetime, and the fluctuations around it will correspond to 
the fluctuations of the physical fields on the spacetime. 

In this paper, I will numerically study the fluctuation spectra around a 
classical solution of a tensor model. This solution has a simple Gaussian form,
and represents a commutative 
but nonassociative fuzzy flat space in arbitrary dimensions \cite{Sasai:2006ua}. 
The main result of this paper is that the number of the low-lying 
low-momentum fluctuation spectra in each momentum sector
agrees exactly with that of a metric tensor modulo the general 
coordinate transformation in the general relativity. 
This suggests the possibility that the effective field theory around the 
classical solution can be described in a similar manner as the general relativity.
If this is definitely proved, this specific tensor model 
contains both the spacetime and
the metric degrees of freedom
in an inseparable fashion, and can be qualified as a quantum gravity.

This paper is organized as follows.
In the following section, a tensor model is defined, which has a one-parameter
family of a Gaussian classical solution.
In Section \ref{sec:procedure}, 
the explicit procedure to obtain the fluctuation spectra and
the zero modes generated from a symmetry 
breaking by a classical solution are discussed.
In Section \ref{sec:flcnum}, discretization of momenta and momentum cutoff are 
introduced as regularization for numerical study, 
and numerical solutions very similar to the Gaussian solution are obtained
in one and two dimensions. The numerical analysis of the fluctuation spectra 
around the numerical solutions is carried out. 
It is shown that there exist some non-zero but very light modes, 
the distribution of which has a characteristic feature depending on the dimensions.   
In Section \ref{sec:distribution}, the distribution of the very light modes 
is studied further by an approximate but more efficient numerical method 
in one to four dimensions. 
In Section \ref{sec:interpretation}, 
the characteristic feature of the distribution of the very light modes 
is shown to be explained in a simple manner, if 
they are identified as the metric in the general relativity.
The final section is devoted to the discussions and future prospects.
 
\section{A tensor model with a Gaussian solution}
The tensor model proposed in \cite{Sasakura:2005js}
has a real three-tensor $C_{abc}$ and a real symmetric 
two-tensor $g^{ab}$ as its dynamical variables.
The model has an invariance under a general linear transformation,
which can naturally be identified as a fuzzy analog of the general 
coordinate invariance in general relativity
under the interpretation of the tensor model as a dynamical theory of fuzzy spaces.
Without any simplifications, 
however, the model is quite complicated in general. 
Therefore, in this paper, it is 
assumed that the two-tensor $g^{ab}$ be non-dynamical, 
and also that the three-tensor $C_{abc}$ be symmetric under the permutations
of the indices.
The former assumption breaks partially the general linear symmetry, 
and the model given below has only an orthogonal group symmetry.
Thus the assumption seems to break the above connection of symmetry 
between the tensor model and the general relativity, 
but the result of this paper will suggest that they are 
actually intimately related. The reason is not clear presently, but it may be
that the general relativity is related to a broader class of tensor models
than expected solely from symmetry. 

The purpose of this section is to construct a tensor model which has a 
classical solution with a simple Gaussian form. This classical solution represents a 
fuzzy space which is essentially the same one introduced in \cite{Sasai:2006ua} 
to study a field theory on a nonassociative spacetime.
In the momentum basis, the fuzzy space is defined by the algebra $f_{p_1}f_{p_2}
={{\bar C}_{p_1p_2}}\,^{p_3}f_{p_3}$ with
\bea
\label{eq:cpgp}
\bar C_{p_1p_2p_3}&=&A \exp \left( -\alpha \left( p_1^2+p_2^2+p_3^2\right) \right)
\delta^D \left( p_1+p_2+p_3 \right), \cr
g^{p_1p_2}&=&\delta^D\left( p_1+p_2 \right),
\eea  
where $A$ and $\alpha$ are positive constants, $D$ is the dimension of the space, and
$p_i$s are continuous $D$ dimensional momentum vectors. 
Only the Euclidean signature is considered in this paper. 
Here I have used the notation $\bar C_{abc}$ with a bar for the Gaussian solution 
to distinguish it from the dynamical variable $C_{abc}$. Similar notations
will be used for some other tensors which will be defined below.
Note that $g^{ab}$ is non-dynamical and takes the fixed value \eq{eq:cpgp}.
In a coordinate basis, after Fourier transformation,  \eq{eq:cpgp} becomes 
\bea
\label{eq:cxgx}
\bar C_{x_1 x_2 x_3}&\sim& \exp \left( -\beta \left((x_1-x_2)^2+(x_2-x_3)^2+(x_3-x_1)^2\right)\right), \cr
g^{x_1x_2}&\sim&\delta^D\left(x_1-x_2\right), 
\eea
where $\beta\sim 1/\alpha$. The reality condition on the dynamical variable
$C_{abc}$ is imposed in this coordinate representation.

The multiplication algebra $f_{x_1}f_{x_2}={\bar C_{x_1x_2}}\,^{x_3}f_{x_3}$ defined by
\eq{eq:cxgx} is commutative but nonassociative. 
This fuzzy space is a fuzzy analogue of a $D$ dimensional flat space.
In fact, \eq{eq:cpgp} (or \eq{eq:cxgx}) has obviously the Poincar\'e symmetry, and, 
in the $\alpha\rightarrow 0\ (\beta\rightarrow \infty)$ limit 
with fixed coordinates\footnote{While the all-over scale has no meaning, since it can be 
rescaled by the rescaling of $\alpha$, the relative scale has a definite meaning.}, 
the multiplication algebra
approaches $f_{x_1}f_{x_2}\sim \delta^D \left(x_1-x_2\right) f_{x_2}$, 
which is the algebra of the functions on a usual
continuum flat space by the identification $f_{x_1}=\delta\left(x-x_1\right)$.

\begin{figure}
\begin{center}
\includegraphics[scale=0.7]{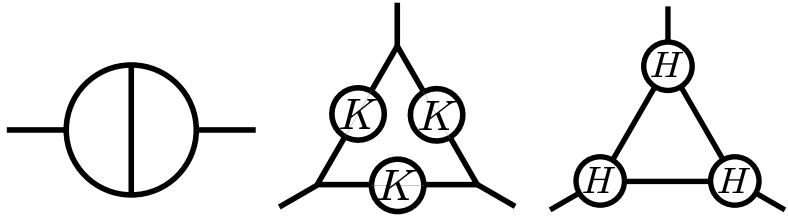}
\end{center}
\caption{From the left to right, 
$K_{ab}$, $H_{abc}$ and $I_{abc}$. The three-vertex represents
the three-tensor $C_{abc}$ and the connecting lines represent the contractions 
with $g^{ab}$. The open ends are associated with the indices of the
tensors. }
\label{fig:khi}
\end{figure}

The nice property of the Gaussian form \eq{eq:cpgp} is that the contraction of the indices
can explicitly be computed, and the resultant tensor has again a Gaussian form. 
The integration measure of momentum in a contraction is defined by 
$\int^\infty_{-\infty} d^D p$. One can easily compute the following quantities 
shown graphically in Figure~\ref{fig:khi}:
\bea
\label{eq:kp}
\bar K_{p_1p_1'}&\equiv&
\bar C_{p_1p_2p_3}\bar C_{p_1'p_2'p_3'}\bar C_{p_4p_4'p_6}\bar C_{p_5p_5'p_6'}
g^{p_2p_4} g^{p_2'p_4'}  g^{p_3p_5}  g^{p_3'p_5'}
g^{p_6p_6'} \cr
&=&
A^4
\left(\frac{\pi}{4\sqrt{2}\alpha}\right)^D \exp\left( -4 \alpha p_1^2 \right) 
\delta^D\left(p_1+p_1'\right)
\eea 
and 
\bea
\label{eq:hp}
\bar H_{p_1p_2p_3}&\equiv &\bar C_{p_1p_4p_5}\bar C_{p_2p_4'p_6} \bar C_{p_3p_5'p_6'}
\bar K^{p_4 p_4'} \bar K^{p_5p_5'} \bar K^{p_6p_6'} \cr
&=&
A^{15}
\left( \frac{\pi}{4 \sqrt{2} \alpha} \right)^{3D} \left( \frac\pi{18\alpha}\right)^{\frac{D}2}
\exp \left(-3 \alpha \left( p_1^2+p_2^2+p_3^2\right)\right) \delta^D\left(p_1+p_2+p_3\right),
\eea
where $\bar K^{p_1p_2}= g^{p_1p_1'} g^{p_2p_2'}\bar K_{p_1'p_2'}$.
Similarly, one finds
\bea
\label{eq:ip}
\bar I_{p_1p_2p_3}&\equiv&\bar H_{p_1p_4p_5}\bar H_{p_2p_4'p_6} \bar H_{p_3p_5'p_6'}
g^{p_4 p_4'} g^{p_5p_5'} g^{p_6p_6'} \cr
&=&
A^{45}
\left(\frac{\pi}{4 \sqrt{2} \alpha}\right)^{9D} \left( \frac{\pi}{18\alpha}\right)^{2D}
\exp \left( -5 \alpha\left(p_1^2+p_2^2+p_3^2 \right)\right)
\delta^D (p_1+p_2+p_3).
\eea

Let us find an equation of motion satisfied by \eq{eq:cpgp}.
From the observation of the decay exponents in \eq{eq:kp} and \eq{eq:ip}, one finds that
the quantity defined by
\be
\label{eq:defofw}
W_{p_1p_2p_3}=C_{p_1p_2p_3}- K^{-1}_{p_1p_1'}K^{-1}_{p_2p_2'}K^{-1}_{p_3p_3'}
I^{p_1'p_2'p_3'},
\ee 
where $K^{-1}_{ab}$ is the inverse of $K^{ab}$,
vanishes for \eq{eq:cpgp}, if $A$ satisfies
\be
1- A^{32} \left(\frac{\pi}{4\sqrt{2}\alpha}\right)^{6D}
\left( \frac{\pi}{18\alpha}\right)^{2D}=0.
\ee
Thus 
\be
\label{eq:wpzero}
W_{p_1p_2p_3}=0
\ee 
is an equation of motion which has \eq{eq:cpgp} as its solution. Note that 
\eq{eq:wpzero} has the solution \eq{eq:cpgp} irrespective of $D$, and 
gives a kind of unified description.

There exist various actions which derive the equation of motion \eq{eq:wpzero}. 
The simplest one would be
\be
\label{eq:action}
S=g^{aa'}g^{bb'}g^{cc'}W_{abc}W_{a'b'c'}.
\ee
Since the action is positive definite and the action vanishes at the solution \eq{eq:cpgp},
the solution is stable, i.e. all the fluctuation spectra are non-negative.

\section{Fluctuation spectra around a classical solution}
\label{sec:procedure}
In this section, I will discuss the explict procedure to obtain the quadratic 
fluctuation spectra of the tensor model \eq{eq:action} around a classical solution,
and also the zero modes generated from a symmetry breaking.

\subsection{The quadratic fluctuation spectra}
The second derivative of $S$ with respect to $C_{abc}$ at a solution $C_{abc}=C^0_{abc}$
to $W_{abc}=0$ is given by 
\be
\left.
\frac{1}{2}\frac{\partial^2 S}{\partial C_{abc} \partial C_{a'b'c'}}\right| _{C=C^0}=
\left.
g^{dd'}g^{ee'}g^{ff'} 
\frac{\partial W_{def}}{\partial C_{abc}}
\frac{\partial W_{d'e'f'}}{\partial C_{a'b'c'}}
\right|_{C=C^0}.
\ee
However, this expression is not appropriate for obtaining the fluctuation spectra.
This is because the derivative should be taken with respect to the independent normalized 
variables, but not all of the components of $C_{abc}$ 
are independent due to the permutation symmetry of its indices. 
In the followings, let me assume that
the range of an index of the tensor is finite, i.e. $a=1,2,\cdots,n$.
This is not true for the continuum momentum index in the preceding section, 
but a regularization will be introduced later for the numerical computations.
Then a natural normalization can be defined by the following $SO(n,R)$ invariant 
measure,
\bea
(\delta C)^2&=&g^{aa'}g^{bb'}g^{cc'} \delta C_{abc} \delta C_{a'b'c'} \cr
&=& \sum_{(p_1,p_2,p_3)} d(p_1,p_2,p_3)\,|\delta C_{(p_1,p_2,p_3)}|^2.
\eea
Here the summation is over the momentum set $(p_1,p_2,p_3)$,
where the order of the momenta $p_1,p_2,p_3$ is not cared, i.e. 
$(p_1,p_2,p_3)=(p_2,p_1,p_3)=\cdots$, and   
$d(p_1,p_2,p_3)$ denotes the number of times the set $(p_1,p_2,p_3)$ appears  
in the simple summation $\sum_{p_1,p_2,p_3}$. Namely,
\be
d(p_1,p_2,p_3)=\left\{ 
\begin{array}{ll}
1 &{\rm for\ }p_1=p_2=p_3, \\
3 &{\rm for\ }p_1=p_2\neq p_3,\ p_2=p_3\neq p_1,\ {\rm or\ }p_3=p_1\neq p_2, \\
6 &{\rm otherwise}.
\end{array}
\right.
\ee 
Therefore the normalized independent variables can be taken as 
\be
\tilde C_{(p_1,p_2,p_3)}=\sqrt{d(p_1,p_2,p_3)}\, C_{(p_1,p_2,p_3)}.
\ee
Thus the fluctuation spectra should be evaluated by
\bea
\label{eq:dsdcdc}
\left.
\frac{1}{2}
\frac{\partial^2 S}{\partial \tilde C_{(p_1,p_2,p_3)} 
\partial \tilde C_{(p_1',p_2',p_3')}}\right| _{C=C^0}
&=&
\sum_{(p_4,p_5,p_6)}
\left.
d(p_4,p_5,p_6) \frac{\partial W_{(p_4,p_5,p_6)}}{\partial \tilde C_{(p_1,p_2,p_3)}}
\frac{\partial W_{(-p_4,-p_5,-p_6)}}{\partial \tilde C_{(p_1',p_2',p_3')}}
\right|_{C=C^0} \cr
&& \hskip-4cm
=
\sum_{(p_4,p_5,p_6)}
\left.
\frac{d(p_4,p_5,p_6)}{\sqrt{d(p_1,p_2,p_3)d(p_1',p_2',p_3')}} 
\frac{\partial W_{(p_4,p_5,p_6)}}{\partial C_{(p_1,p_2,p_3)}}
\frac{\partial W_{(-p_4,-p_5,-p_6)}}{\partial C_{(p_1',p_2',p_3')}}
\right|_{C=C^0} \\
&&\hskip-6cm =
\sum_{(p_4,p_5,p_6),\sigma,\sigma'}
\left.
\frac{d(p_4,p_5,p_6) \sqrt{d(p_1,p_2,p_3)d(p_1',p_2',p_3')}} {3!\ 3!} 
\frac{\partial W_{(p_4,p_5,p_6)}}{\partial C_{p_{\sigma(1)}p_{\sigma(2)}p_{\sigma(3)}}} 
\frac{\partial W_{(-p_4,-p_5,-p_6)}}{\partial C_{p_{\sigma'(1)}'p_{\sigma'(2)}'p_{\sigma'(3)}'}} \right|_{C=C^0}, \nonumber
\eea
where the partial derivatives in the last line are taken as if all the $C_{abc}$
are independent, and the sum over $\sigma,\sigma'$ is 
over all the permutations of $1,2,3$ to compensate this virtual treatment.
The fluctuation spectra can be obtained by diagonalizing the matrix \eq{eq:dsdcdc}.

Since $W_{abc}$ is a complicated function of $C_{abc}$, the 
computation of the explicit formula of the partial derivatives in \eq{eq:dsdcdc} 
is straightforward but lengthy. All the necessary formula are shown in Appendix A.  

\subsection{The zero modes from the symmetry breaking $SO(n,R)/SO(2)^D$}
\label{sec:zeromodes}
In this subsection, I will study the zero modes of the fluctuation spectra. A number of
zero modes appear, since a classical solution breaks the $SO(n,R)$ symmetry of 
the tensor model in general. Let me start with the standard discussion.
 
Suppose there is a classical solution $C_{abc}=C^0_{abc}$,
\be
\label{eq:eomgeneral}
\left.
\frac{\partial S}{\partial C_{abc}}\right|_{C=C^0}=0.
\ee
The $SO(n,R)$ symmetry of the action implies that, a new solution can be generated by
applying a broken transformation to $C^0_{abc}$. By considering 
an infinitesimally small transformation of \eq{eq:eomgeneral}, this implies
\be
\left.
\frac{\partial ^2 S}{\partial C_{abc} \partial C_{def}}\right|_{C=C^0}
({M_d}^{d'}C^0_{d'ef}+{M_e}^{e'}C^0_{de'f}+{M_f}^{f'}C^0_{def'})=0,
\ee
where ${M_a}^{a'}$ is a broken generator. Namely,
\be
\label{eq:zeromode}
\delta C^M_{abc}={M_a}^{a'}C^0_{a'bc}+{M_b}^{b'}C^0_{ab'c}+{M_c}^{c'}C^0_{abc'}
\ee
is a zero mode of the quadratic fluctuation spectra.
Thus, in general, there is a one-to-one correspondence between the broken generators
and the zero modes of the fluctuation spectra around a classical solution.

The general coordinate transformation of the metric tensor, 
$\delta g_{\mu\nu}=\nabla_\mu v_\nu+\nabla_\nu v_\mu$, 
in the general relativity
is not linear in $g_{\mu\nu}$. There is an old idea that 
the non-linear transformation comes
from a spontaneous breaking of the local translational symmetry \cite{Borisov:1974bn},
in analogy with the standard non-linear realizations of global symmetries.
The present setting of the tensor model fits very well with this old idea, since
the symmetry of the tensor model can naturally be
identified with the fuzzy analogue of the general coordinate invariance
\cite{Sasakura:2005js} and 
a classical solution breaks the symmetry in general. 
Then the zero modes of the fluctuation spectra should be 
identified with these gauge zero modes, which are unphysical.

In the numerical investigations in the following sections, 
the values of momenta will be discretized in integers, and a cutoff will be  
introduced. Namely, an index of the tensor takes the form,
\be
\label{eq:a}
a=(p^1,p^2,\cdots,p^D),\ \ (p^i=-L,-L+1,\cdots,L).
\ee
I will also make an ansatz that there remains a $D$-dimensional translational invariance $SO(2)^D$ on a solution.
These assumptions mean that the background space corresponding to a solution 
is actually a fuzzy $D$-dimensional flat torus. 
The broken symmetry is given by $SO(n,R)/SO(2)^D$, and thus 
the total number of the zero modes is given by
\be
\label{eq:totalzero}
\# {\rm zero}_{total}=\frac{n(n-1)}{2}-D,
\ee
where $n$ is the size of the range of the index \eq{eq:a}, i.e. $n=(2L+1)^D$.

The above counting of the zero modes can be made more detailed by counting them
in each momentum sector, which will be used 
in numerical investigations in the following sections.
The generator of $SO(2)^D$ is given by 
\be
\label{eq:unbroken}
M^{a}_{p_1p_2}= i (v \cdot p_1) \delta_{p_1\ -p_2},
\ee
where $v$ is a real $D$-dimensional vector.
It acts on $C_{abc}$ as
\be
i\, v\cdot (p_1+p_2+p_3)\, C_{p_1p_2p_3}.
\ee 
Therefore, because of the translational symmetry, 
the tensor of a solution $C^0_{p_1p_2p_3}$ takes non-vanishing values only 
at the vanishing momentum, $p_1+p_2+p_3=\vec 0$.
This implies that the momentum of a zero mode $\delta C^M_{abc}$ in \eq{eq:zeromode}
comes solely from that of a generator $M_{ab}$.
Thus counting the zero modes in each momentum sector 
is equivalent to counting the broken generators in each momentum sector.

There are two conditions for $M_{ab}$ to be a generator of $SO(n,R)$ as
\bea
\label{eq:condm}
M_{p_1p_2}^*&=&M_{-p_1\,-p_2}, \cr
M_{p_1p_2}&=&-M_{p_2p_1},
\eea
where $*$ denotes the complex conjugate. The first condition comes from the reality of the 
generator in the coordinate representation\footnote{Note that the reality condition 
is imposed in the coordinate representation like 
\eq{eq:cxgx}.}, and the second one comes 
form the anti-symmetric property of the indices of the generators of $SO(n,R)$. 
The total momentum of 
$M_{p_1p_2}$ is $p=p_1+p_2$. 

When the total momentum vanishes, i.e. $p=\vec 0$, the conditions \eq{eq:condm}
are given by
\bea
\label{eq:condmzero}
M_{-p_1\, p_1}^*&=&M_{p_1\,-p_1}, \cr
M_{-p_1\, p_1}&=&-M_{p_1\, -p_1},
\eea
where $p_1$ is an integer vector as \eq{eq:a}.
These conditions imply that $M_{\vec 0,\vec 0}$ must vanish, and for $p_1\neq \vec 0$, $M_{p_1\,-p_1}$
are pure imaginary and only half of the generators are independent.
Therefore the number of the independent generators in the $p=\vec 0$ sector is given by
\be
\label{eq:nummzero}
\# M (\vec 0 ) =\frac12 (n-1).
\ee

When the total momentum $p$ does not vanish, the conditions become
\bea
\label{eq:condmnotzero}
M_{-p_1+p\ p_1}^*&=&M_{p_1-p\ -p_1}, \\
M_{-p_1+p\ p_1}&=&-M_{p_1\ -p_1+p},
\eea
where both $p_1$ and $-p_1+p$ must be integer vectors as \eq{eq:a}.
The second condition implies that, when there exists a $p_1$ which satisfies $2p_1=p$, i.e.
when $p$ is an even vector, $M_{p_1p_1}$ must vanish. It also implies that,
for $2 p_1\neq p$, only half of $M_{-p_1+p\ p_1}$ are independent, each of which is 
a complex number.
The first condition implies that the generators in the momentum $-p$ sector is determined
by those in the momentum $p$ sector. In fact, under the reality condition in the coordinate 
basis, the modes must be a combination of those in the $p$ and $-p$ sectors, in the same 
was as $\sin(px)=(e^{ipx}-e^{-ipx})/(2i)$ or $\cos(px)=(e^{ipx}+e^{-ipx})/2$.
Therefore this effectively generates an additional factor $1/2$ to each $p$ and $-p$ 
sector\footnote{This rough 
treatment can be justified in the actual usage in the following numerical study.}. 
Counting the number of the possible $p_1$ 
under the condition that both $p_1$ and $-p_1+p$ must be 
in the range of \eq{eq:a}, and subtracting 1 when $p$ is an even vector, one finally obtains
the number of the independent generators of $SO(n,R)$ in the momentum $p$ sector as 
\be
\label{eq:nummgen}
\# M(p)=\frac12 \left( \prod_{i=1}^D \left( 2L+1-|p^i|\right) -{\rm even}(p)\right),
\ee
where $p^i$ is the $i$-th component of the momentum $p$, and 
\be
{\rm even}(p)=\left\{
\begin{array}{cl}
1 & {\rm for\ }p \ {\rm  even\ vector}, \\
0 & {\rm otherwise}.
\end{array}
\right.
\ee
Note that the result \eq{eq:nummgen} includes also the special case \eq{eq:nummzero}. Note also
that, since the momentum of an index is bounded as \eq{eq:a}, 
each component $p^i$ of the total momentum $p=p_1+p_2$ in \eq{eq:nummgen} is 
bounded by $|p^i|\leq 2L$. 

The unbroken genenerators \eq{eq:unbroken} are in the $p=\vec 0$ sector. Subtracting 
this from \eq{eq:nummgen}, one obtains the formula for the number of the zero modes 
in each momentum sector as
\be
\label{eq:numzero}
\# {\rm zero}(p)=
\left\{ 
\begin{array}{ll}
\frac12 \left( \prod_{i=1}^D \left( 2L+1-|p^i|\right) -{\rm even}(p)\right)
-D \delta_{p\vec 0} &{\rm for\ } |p^i|\leq 2L,\\
0&{\rm otherwise}. 
\end{array}
\right.
\ee
    
\section{Fluctuation spectra around the Gaussian-like 
numerical solutions in one and two dimensions}
\label{sec:flcnum}
The Gaussian solution \eq{eq:cpgp} is a function of continuous infinite momenta, 
and is not suited for direct numerical investigations. 
Therefore \eq{eq:a} is introduced as regularization.  Then, however,
\eq{eq:cpgp} does not satisfy the equation of motion \eq{eq:wpzero}.
There are two ways to deal with this problem. One is to obtain numerical solutions 
similar to \eq{eq:cpgp} to the equation of motion \eq{eq:wpzero}, and study numerically 
the fluctuation spectra around it.
The other is to use \eq{eq:cpgp} as an approximate solution to the equation of motion 
\eq{eq:wpzero}, 
and evaluate numerically the spectra by substituting \eq{eq:cpgp} directly 
into \eq{eq:dsdcdc}.
The former way will be carried out in this section, and the latter will be 
the subject in the next section.
The former way is definite but takes longer computational time, 
and will be carried out only in
dimensions $D$=1,2 with smaller $L$.
The latter way is an approximate treatment but is more efficient, 
since it does not contain the process of obtaining numerical solutions, and will be carried out in $D=1,2,3,4$ for larger $L$.

The numerical computation is carried out on a Windows XP 64-bit  
workstation with two AMD Opteron 275 processors (dual core, 2.2GHz). 
The total memory size is 8 GB.
The main numerical computation is carried out with Intel C++ compiler 9.1 
and 10.0 with Open MP parallelization. 
AMD core math library (ACML) 3.6.0 is used for some basic routines.
Some supportive computations such as drawing graphs and cross checks are carried out with Mathematica 5.2 personal grid edition.

Let me first consider the $D=1$ case. 
Let me make some ansatz in obtaining the solutions to 
\eq{eq:wpzero}.  
One is the translational invariance of the solution, namely, 
$C^0_{p_1p_2p_3} \neq 0$ only at $p_1+p_2+p_3=0$. 
This was assumed in counting the zero modes in Section~\ref{sec:zeromodes}.
Another is that $C^0_{p_1p_2p_3}$ be real.
The last one is
\be
\label{eq:refd1}
C^0_{p_1p_2p_3}=C^0_{-p_1\, -p_2\, -p_3},
\ee
which is a reflection symmetry of the corresponding 
fuzzy $S^1$.

The numerical solutions are obtained by taking starting values of $C_{p_1p_2p_3}$
as \eq{eq:cpgp} with $\alpha\sim1/L^2$, 
searching for the minimum of the sum of the equation of motion \eq{eq:wpzero},
\be
\label{eq:fc}
f(C)=
\sum_{(p_1p_2p_3)} W_{p_1p_2p_3}^2,
\ee 
and checking whether the minimum can reasonably be regarded as vanishing
within the computational accuracy.
The Nelder-Mead method is used for the optimization.

The numerical computation is carried out for $L=3,5,10$.
The raw numerical results are shown in Appendix \ref{ap:B}. 
In Figure~\ref{fig:Lk}, the values of $K_p^p$, which is defined by neglecting the bar
$\bar\ $
in \eq{eq:kp} (or in Figure~\ref{fig:khi}), are plotted for the numerical solutions.  
The shapes look very much like Gaussian, and therefore 
the numerical solutions can be regarded as the correct
discrete solutions to \eq{eq:wpzero} corresponding to the continuum Gaussian solution 
\eq{eq:cpgp}.   
\begin{figure}
\begin{center}
\includegraphics[scale=.6]{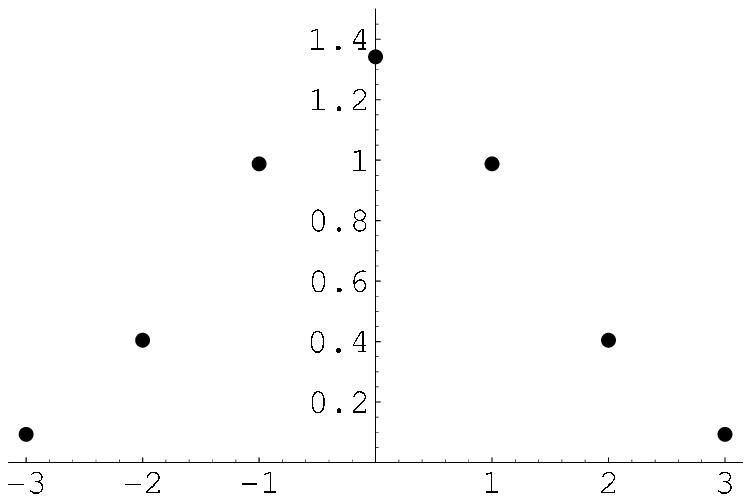}
\includegraphics[scale=.6]{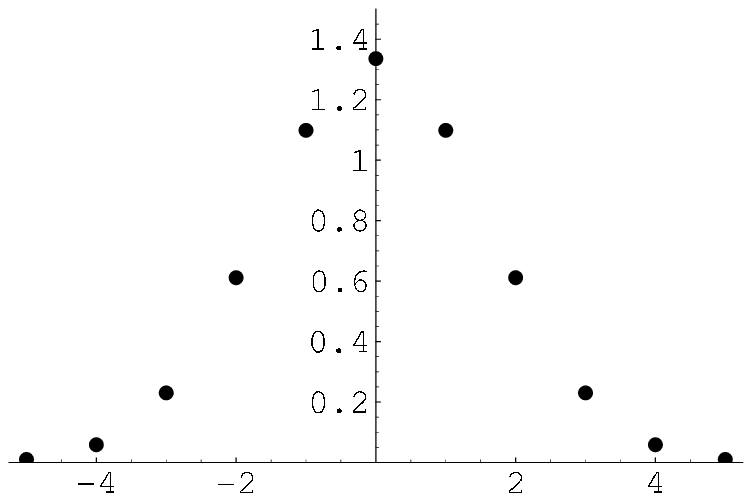}
\includegraphics[scale=.6]{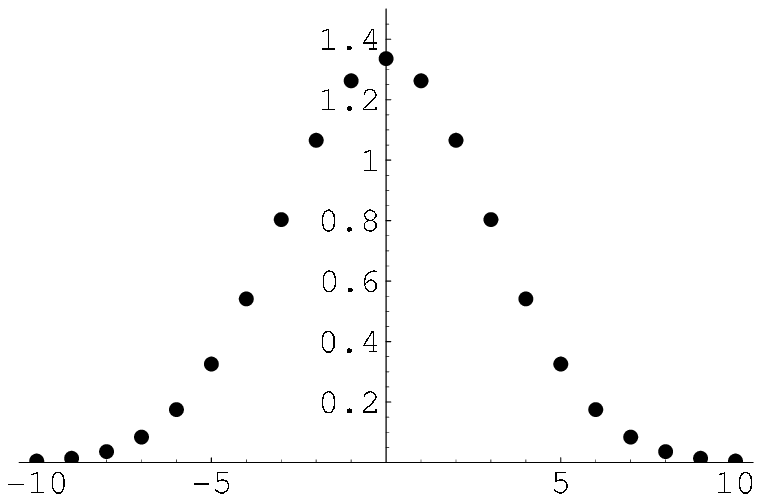}
\end{center}
\caption{The values of $K_p^p$ for $D=1$, $L=3,5,10$ (left to right). 
The horizontal axis is $p$.  }
\label{fig:Lk}
\end{figure}

\begin{figure}
\begin{center}
\includegraphics[scale=.6]{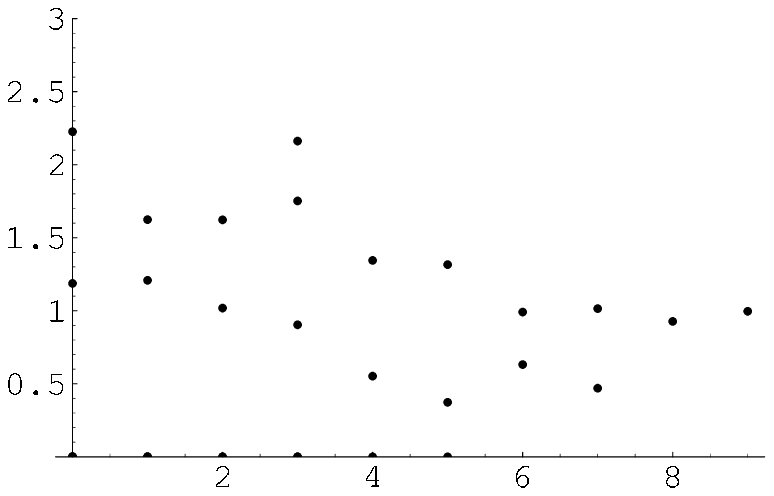}
\includegraphics[scale=.6]{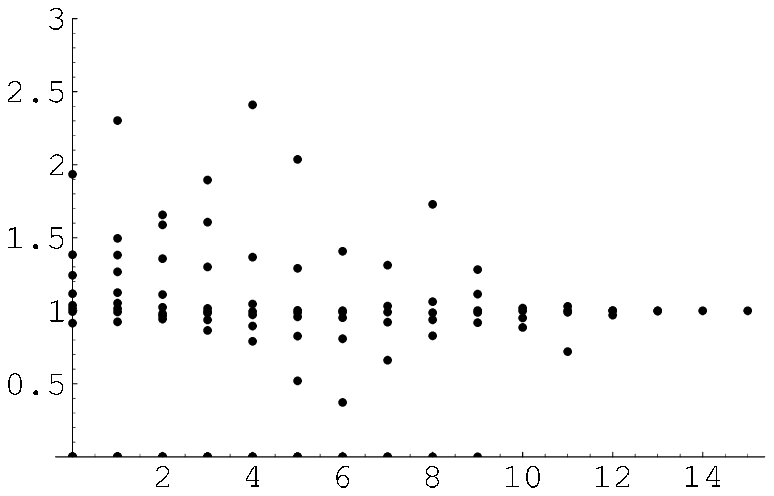}
\includegraphics[scale=.6]{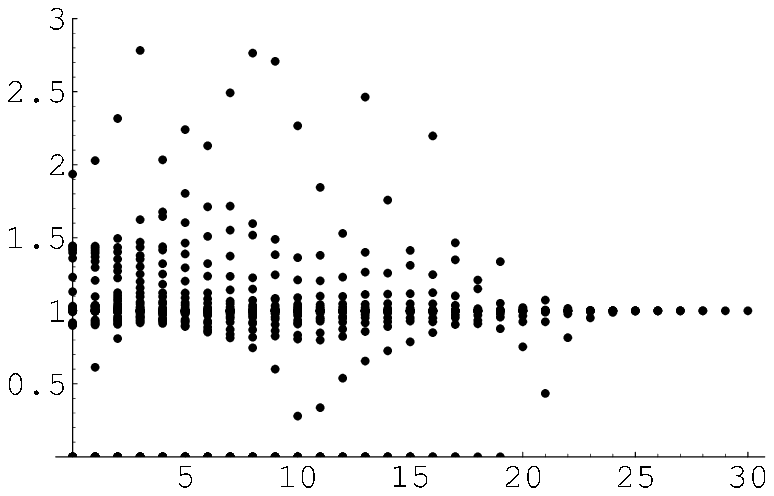}
\end{center}
\caption{The fluctuation spectra for $D=1$, $L=3,5,10$ (left to right). 
The horizontal axis is the momentum 
$p$, and the vertical one is the spectrum.}
\label{fig:Lspec}
\end{figure}

Let me first look at the $L=3$ case. 
The spectra in Table~\ref{tab:L3} contain values $\lesssim 10^{-14}$. The number of these tiny spectra in each momentum 
sector agrees with that of the zero modes from the formula \eq{eq:numzero}. Therefore these
modes can be identified with the zero modes generated by the symmetry 
breaking $SO(n,R)/SO(2)$.
Another thing one notices is that there exists a non-zero but very light mode 
with $\sim 2\times 10^{-3}$ 
only at the $p=0$ sector, while the others are $\gtrsim O(0.1)$. 
The qualitative features of the spectra for $L=5,10$ are similar with $L=3$, i.e. 
the agreement with \eq{eq:numzero}, the existence of a very light mode only at
the $p=0$ sector,
and that the others are $\gtrsim O(0.1)$. The very light modes have the spectra
$\sim 1\times 10^{-6}$ and $\sim 3\times 10^{-7}$ for $L=5,10$, respectively. 
The spectra are plotted in Figure~\ref{fig:Lspec}. 

The $D=2$ case is carried out essentially in a similar manner. To reduce the 
number of the variables in searching for a solution, some reflection 
symmetries are imposed.
One is the reflection symmetry in each direction,
\be
C^0_{(p^1_1,p^2_1)\, (p_2^1,p_2^2)\, (p_3^1,p_3^2)}=
C^0_{(-p^1_1,p^2_1)\, (-p_2^1,p_2^2)\, (-p_3^1,p_3^2)}=
C^0_{(p^1_1,-p^2_1)\, (p_2^1,-p_2^2)\, (p_3^1,-p_3^2)}=
C^0_{(-p^1_1,-p^2_1)\, (-p_2^1,-p_2^2)\, (-p_3^1,-p_3^2)}.
\ee  
The other is a symmetry which exchanges the two coordinates, 
\be
C^0_{(p^1_1,p^2_1)\, (p_2^1,p_2^2)\, (p_3^1,p_3^2)}=
C^0_{(p^2_1,p^1_1)\, (p_2^2,p_2^1)\, (p_3^2,p_3^1)}.
\ee
The meaning of these symmetries is that the space is a fuzzy analogue of 
$S^1\times S^1$, where the two $S^1$'s are congruent and 
perpendicular to each other, and each of them has a reflection
symmetry.

\begin{figure}
\begin{center}
\includegraphics[scale=.7]{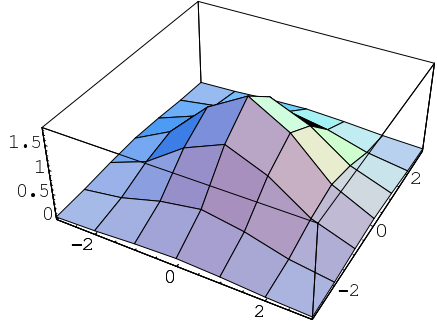}
\end{center}
\caption{The profile of $K_{(p^1,p^2)}^{(p^1,p^2)}$ for $D=2,L=3$. The horizontal axes are $(p^1,p^2)$.}
\label{fig:d2l3k}
\end{figure}

\begin{figure}
\begin{center}
\includegraphics[scale=1]{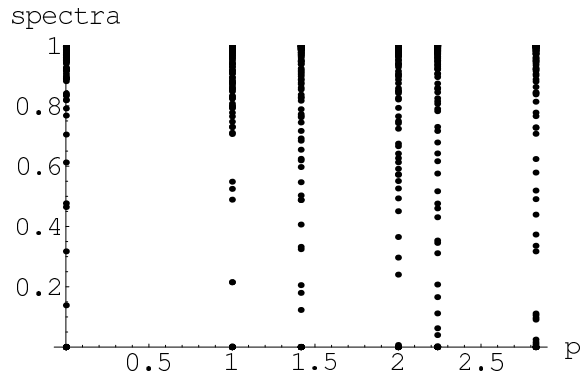}
\end{center}
\caption{The fluctuation spectra for $D=2, L=3$.}
\label{fig:specd2l3}
\end{figure}

The numerical computation is carried out for $L=2,3$. 
Numerical solutions similar to \eq{eq:cpgp} are found. In fact, $K_{(p^1,p^2)}^{(p^1,p^2)}$ for the numerical solution for $L=3$ is plotted  
in Figure~\ref{fig:d2l3k}, the profile of which is similar to Gaussian.
The raw numerical results are summarized in Appendix~\ref{ap:B}. 
The spectra at low momentum sectors for $L=3$ are plotted in Figure~\ref{fig:specd2l3}.
Some qualitative features are similar to the $D=1$ case.  
The tiny spectra can be identified with the zero modes, 
the number of which agree with \eq{eq:numzero}.
There exist also non-zero but very light modes, 
the distribution of which, however, is different from the $D=1$ case.
There exist three very light modes at the $p=(0,0)$ sector, and one very 
light mode at each non-vanishing momentum sector. 
For $L=3$, the spectra of these very light modes are $1.3\times 10^{-12}$, $3.0\times 10^{-4}$ and $1.1\times 10^{-3}$ at the $p=(0,0)$ sector, 
and $2.1 \times 10^{-3}$, $9.6\times 10^{-4}$ and 
$6.9\times 10^{-3}$ in the $p=(1,0),(1,1)$ and $(2,0)$, respectively.  
These modes are clearly distinguished from the other heavy modes with
spectra $\gtrsim O(0.1)$.
In the larger momentum sectors, however, the hierarchy between the very light modes
and the other heavier modes cannot be clearly observed.
This situation is expected to be improved in larger $L$.

The number of the dynamical variables $C_{abc}$ expands very rapidly with
the increase of $D,L$, and it seems very hard
to carry out the same computations for larger $D,L$ in the present 
computational facilities. Our final goal is of course $D=4$. Therefore,
in the next subsection, approximate treatment will be carried out for larger 
$D,L$ including $D=4$.
The numerically correct results in this section will be used to support the results 
of the approximate treatment in the following section. 

\section{Distribution of the very light modes with an approximate method}
\label{sec:distribution}
The results in the previous section show that the spectra can be classified into
three categories.
One consists of the heavy modes with spectra $\gtrsim O(0.1)$,
another consists of the very light modes, and the other
consists of the zero modes from the symmetry breaking $SO(n,R)/SO(2)^D$.
The distribution of the very light modes seems to be characteristic of the 
dimensions $D$. 
The main purpose of the present section is to study numerically the distribution 
of the very light modes in one to four dimensions by using a more efficient but 
approximate method. 
As has been checked for some $D,L$ in the preceding section, 
I assume that the number of vanishing
spectra is given by \eq{eq:numzero} also for other $D,L$. 
Namely, zero modes are in one-to-one correspondence to the broken generators of
the symmetry breaking $SO(n,R)/SO(2)^D$. 

An inefficient part of the numerical computations in the previous section is 
the process to find the numerical solutions by minimizing $f(C)$. 
The reasons for the inefficiency are not only 
the complexity of the equation of motion 
\eq{eq:wpzero}, but also the existence of the very small spectra at the 
$p=0$ sector. The function $f(C)$ has a very gentle slope in this direction,
and it is hard to find its minimum.
This time-consuming step can be skipped, 
if the analytical expression \eq{eq:cpgp} can be used as an approximation to a
solution to the equation of motion \eq{eq:wpzero}.  
The errors of the approximation come from the cutoff and the discretization of 
momentum introduced in \eq{eq:a}.
One may appropriately choose the parameter $\alpha$ of the analytic solution \eq{eq:cpgp}
so that the approximation become accurate enough to study the 
distribution of the very light modes.

If the value of $\bar C_{abc}$ in \eq{eq:cpgp} 
is small outside the range of \eq{eq:a}, 
the introduction of the cutoff will cause small errors. This requires
\be
\label{eq:alcutoff}
\alpha L^2 \gtrsim 1.
\ee  
On the other hand, if $\tilde C_{abc}$ changes gently as a function of momenta,
the introduction of discretization will cause small errors. This requires
\be
\label{eq:aldist}
\alpha (p^2-p'^2) \lesssim 1 
\ee
for nearby $p,p'$.
Combining the two conditions, one obtains
\be
\frac1{L^2} \lesssim \alpha \lesssim \frac{1}{L}.
\ee
This inequality suggests that the continuum limit $L\rightarrow \infty$ 
can safely be taken by simultaneously tuning $\alpha$ in the allowed region.

The actual process of computing the fluctuation spectra is as follows. 
The fluctuation matrix \eq{eq:dsdcdc} is a product of two matrices
$\left. \frac{\partial W_{abc}}{\partial C_{def}}\right|_{C=\bar C}$. 
As shown in Figure~\ref{fig:km1} and \ref{fig:idc}, 
$\left.\frac{\partial W_{abc}}{\partial C_{def}}\right|_{C=\bar C} $ contains some loops. 
The contribution of these loops can be analytically evaluated by Gaussian
integrations, if 
the integration variables are the unbounded continuum momenta,
but not if they are the bounded discrete momenta \eq{eq:a}.
In the latter case, the loop integrals must be treated numerically. 
Moreover the existence of the momentum cutoff would also require numerical treatment for the internal tree lines.
Therefore, to avoid these complexities,
I employ the unbounded continuous momenta for internal lines, and 
compute analytically 
$\left. \frac{\partial W_{abc}}{\partial C_{def}}\right|_{C=\bar C}$.
And then the external momenta, i.e. the indices of $\left. \frac{\partial W_{abc}}{\partial C_{def}}\right|_{C=\bar C}$, 
are assumed to take the bounded discretized  momenta \eq{eq:a},
and the fluctuation matrix \eq{eq:dsdcdc} is diagonalized numerically.
This procedure would be allowed as an approximation.
The analytical expressions of the results of the loop integrals in
$\left. \frac{\partial W_{abc}}{\partial C_{def}}\right|_{C=\bar C}$ 
are given in Appendix~\ref{app:gauss}.

\begin{figure}
\begin{center}
\includegraphics[scale=1.]{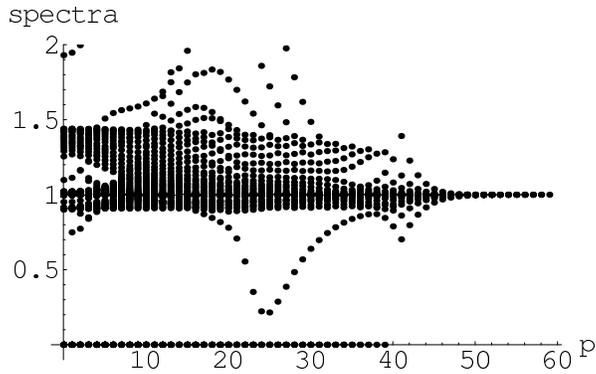}
\end{center}
\caption{The spectra for $D=1,L=20,\alpha=\frac{1}{L^2}$. }
\label{fig:specd1l20}
\end{figure}

Let me first show the result of the numerical computation in $D=1,L=20$,
with the choice of the parameter $\alpha=\frac{1}{L^2}$.
The spectra are plotted in Figure~\ref{fig:specd1l20}.
There one can see the clear hierarchy between the modes with the ``vanishing" spectra 
within the resolution of the figure and the other ``heavy" modes with
the spectra $\gtrsim O(0.1)$.
The number counting of the ``vanishing" spectra
reproduces the result in the preceding section. At $p=0$ sector, 
the number is larger by one than
the formula \eq{eq:numzero}, and agrees with \eq{eq:numzero} at $p\neq 0$ sectors.
This implies that there exists a very light mode only at the $p=0$ sector.

Unfortunately, this approximate treatment is not good enough 
to extract the very light mode. 
In fact, the largest value of the ``vanishing" spectra at the $p=0$ sector is 
$\sim3\times 10^{-5}$, while the second largest is $\sim 7\times 10^{-6}$,
and no clear hierarchy between the zero modes and the very light mode 
can be observed.
Note that, in the correct numerical computation in the previous section,  
there existed a clear hierarchy, e.g. $\sim 10^{-6}$ and $\sim 10^{-15}$ 
for $D=1,L=5$.
Therefore the approximate treatment in this section would be 
enough only to distinguish between
the ``heavy" modes and the bunch of the zero and the very light modes.
However, this would be enough to study the distribution of the very light modes 
by counting the number of the bunch of ``vanishing" spectra 
and subtracting the number of 
the zero modes \eq{eq:numzero}.

\begin{figure}
\begin{center}
\includegraphics[scale=.5]{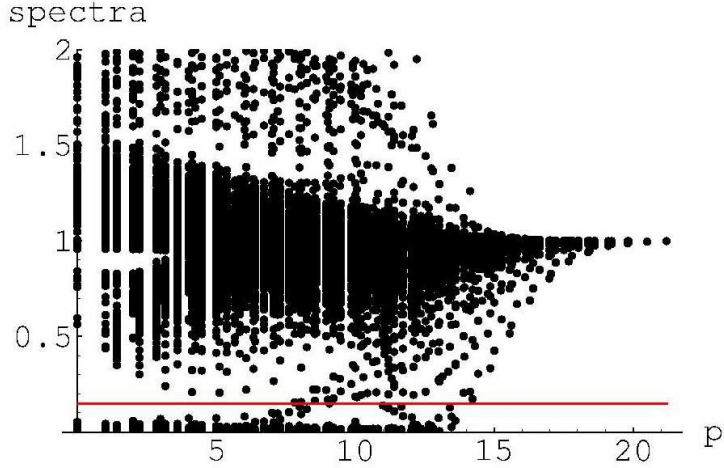}
\end{center}
\caption{The spectra for $D=2,L=5,\alpha=\frac{0.2}{L^2}$. The horizontal axis is 
$p=\sqrt{(p^1)^2+(p^2)^2}$. The horizontal line is drawn at $0.15$.}
\label{fig:specd2l5al02}
\end{figure}

The spectra for $D=2,L=5,\alpha=\frac{0.2}{L^2}$ are plotted in Figure~\ref{fig:specd2l5al02}.
There one finds 
that the ``vanishing" spectra form a finite band from the origin of the vertical axis,
which would be due to the approximate treatment.
A hierarchy between the ``vanishing" spectra and the other ``heavy" ones can
be clearly observed at $p=\sqrt{(p^1)^2+(p^2)^2}\lesssim 7$.
One can count the numbers of the ``vanishing" modes by counting the numbers of modes 
below a value, say $\lesssim 0.15$. This reproduces the observation made in the 
previous section. Namely, the excess of the number in comparison with the formula 
\eq{eq:numzero} is three at $p=0$, and one at $p\neq 0$. 

In the figure, one notices that there exist a series of spectra which leave
the ``vanishing" spectra at $p\sim 7$ and gradually merges to the ``heavy" modes at 
$p \sim 12$. This kind of behavior does not exist in the $D=1$ case. These facts 
suggest that the very light mode begins to mix with the ``heavy"
modes at $p \gtrsim 7$.

\begin{figure}
\begin{center}
\includegraphics[scale=.27]{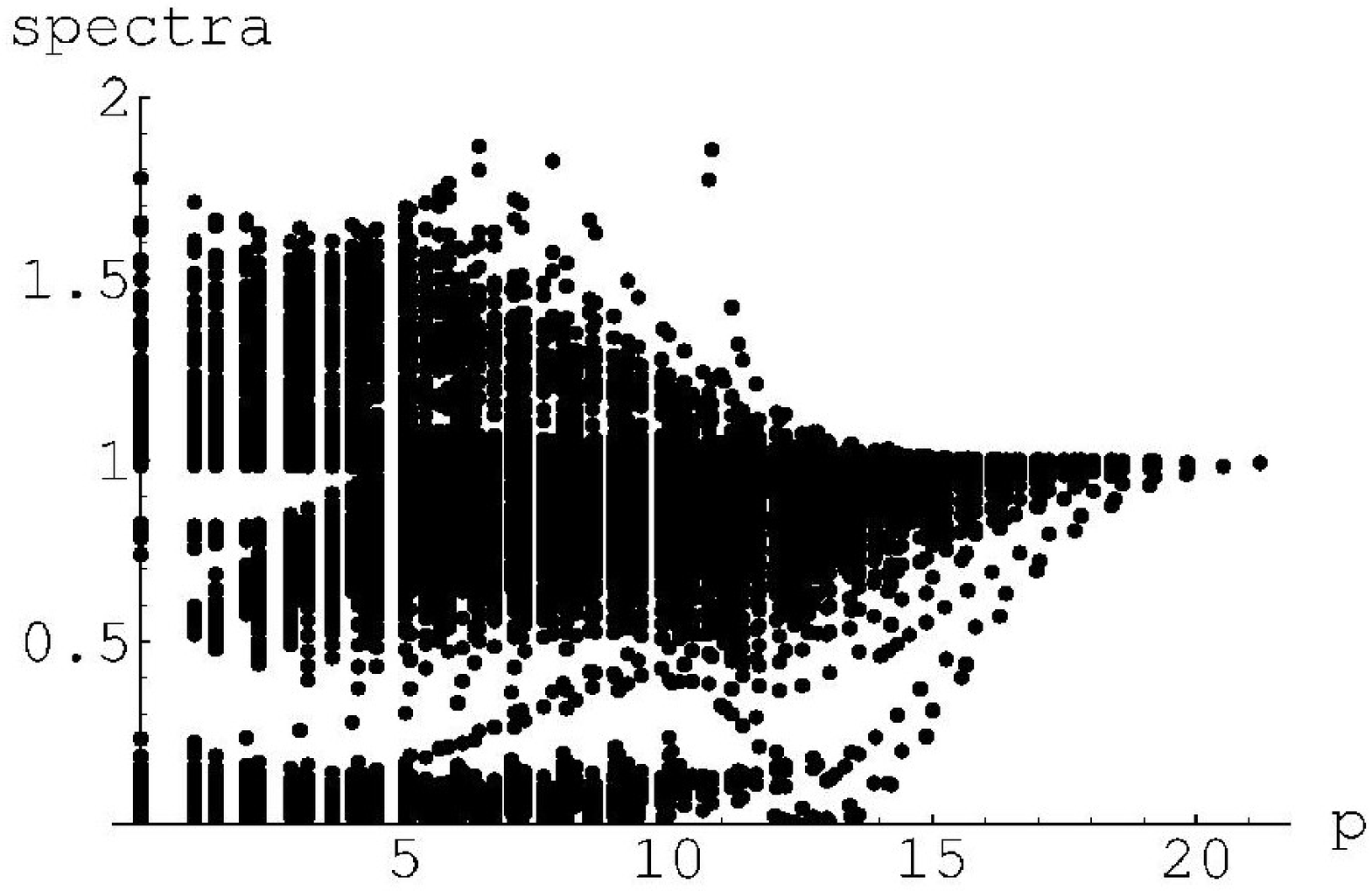}
\includegraphics[scale=.27]{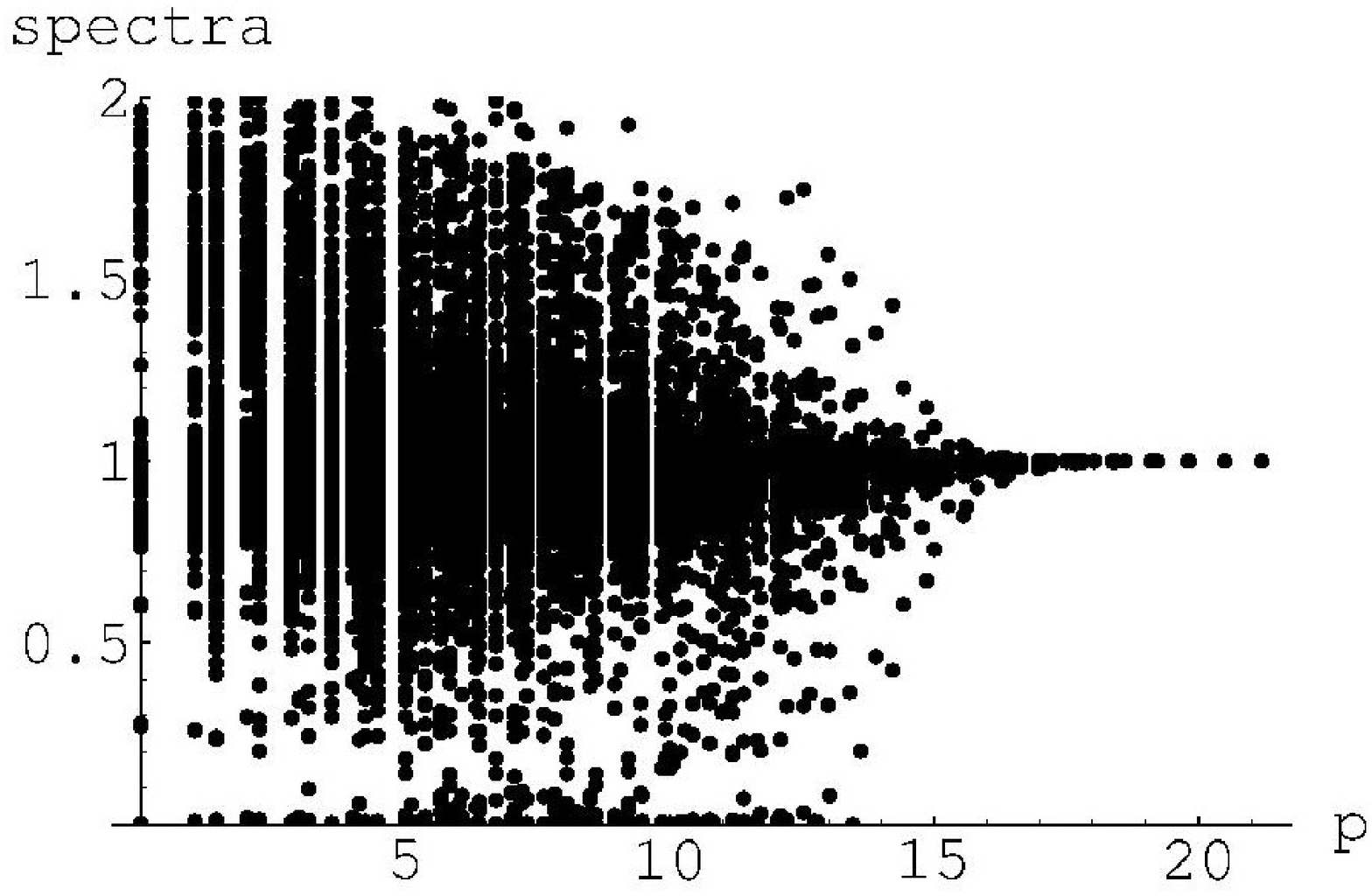}
\includegraphics[scale=.27]{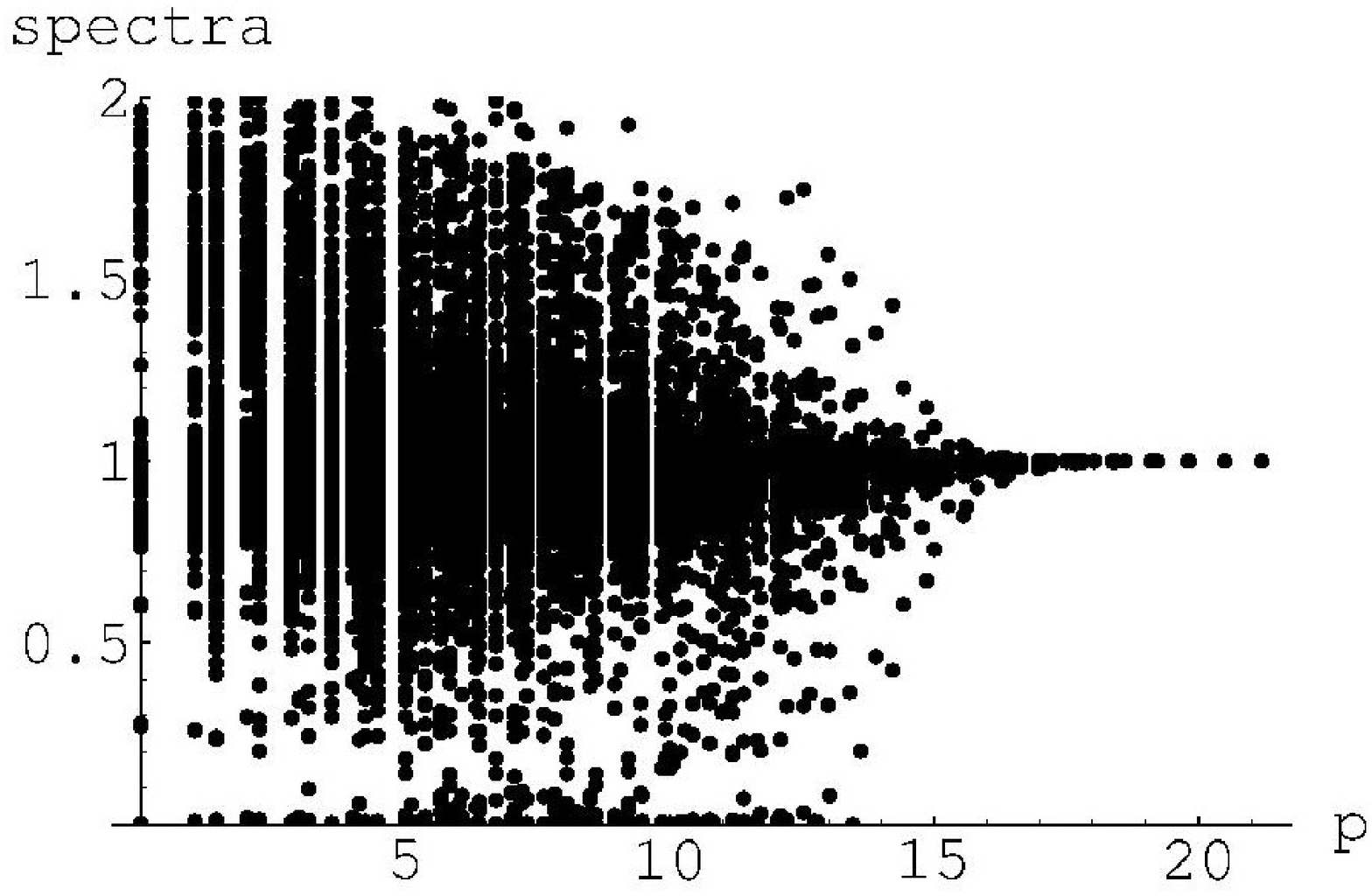}
\end{center}
\caption{The results for $D=2,L=5,\alpha=\frac{0.1}{L^2}$ (left), 
$D=2,L=5,\alpha=\frac{0.5}{L^2}$ (middle) 
and $D=2,L=10,\alpha=\frac{0.2}{L^2}$ (right). 
The last case is studied only for the momentum sectors $(p^1,p^2)\ |p^i|\leq 6$, 
because of the long computational time. }
\label{fig:specd2l5al05}
\end{figure}

The fluctuation spectra for some other parameters are shown in Figure~\ref{fig:specd2l5al05}.
The hierarchies between the bunch of spectra near the origin and the others can
be observed more or less clearly in each figure.
One can see that, as $\alpha$ becomes smaller, 
the width of the bunch of the ``vanishing" spectra tends to become larger, and, 
as $\alpha$ becomes larger, the ``vanishing" and  ``heavy" spectra tend to merge
at lower momentum sectors.  
Therefore, to obtain a clear hierarchy between the ``vanishing" and the ``heavy"
modes in a wide range of momentum sectors, 
the parameter $\alpha$ must be chosen appropriately. 
An appropriate continuum limit $L\rightarrow\infty$ will be obtained with
this kind of tuning of $\alpha$.  
Counting the number of the spectra in each bunch from the origin 
at low momentum sectors, 
the same distribution of the very light modes as above is obtained.

\begin{figure}
\begin{center}
\includegraphics[scale=1]{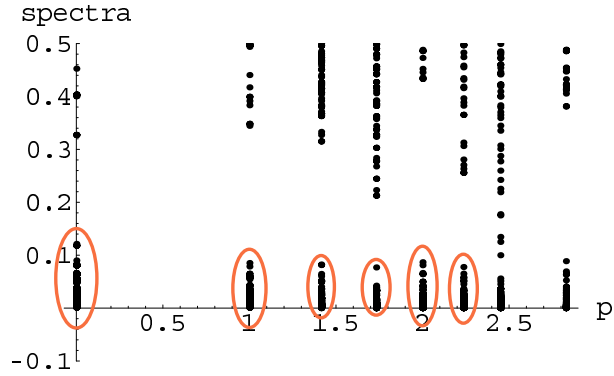}
\end{center}
\caption{The fluctuation spectra for $D=3,L=3,\alpha=\frac{0.15}{L^2}$.
The numbers of the spectra near the origin surrounded by the circles are counted and
compared with the formula \eq{eq:numzero}. }
\label{fig:specd3l3al015}
\end{figure}

In the followings, let us study the higher dimensional cases.
In Figure~\ref{fig:specd3l3al015}, the fluctuation spectra for $D=3,L=3,\alpha=\frac{0.15}{L^2}$ are shown. 
A clear gap between the bunch of spectra from the origin 
and the others can be observed at 
the momentum sectors $p<2.4$. 
By counting the number of the spectra surrounded by the circles in the figure,
the excess over the formula \eq{eq:numzero} is
obtained as 6 at $p=0$, and 3 at $p \neq 0$.

\begin{figure}
\begin{center}
\includegraphics[scale=1]{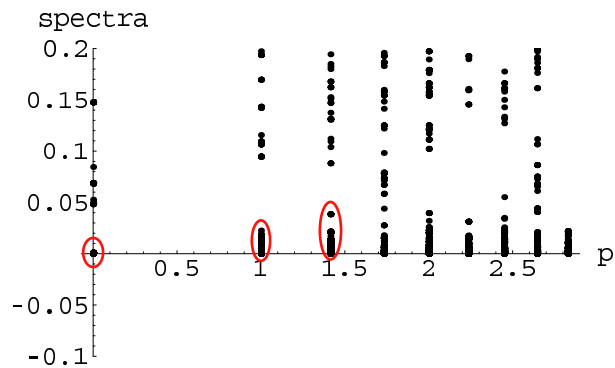}
\end{center}
\caption{The fluctuation spectra for $D=4,L=2,\alpha=\frac{0.3}{L^2}$. With this small
value of $L$, the hierarchy between the bunch of spectra near the origin (the spectra surrounded by circles)
and the others is clear only for $p=0,1,\sqrt{2}$. }
\label{fig:specd4l2al03}
\end{figure}

In Figure~\ref{fig:specd4l2al03}, the fluctuation spectra for $D=4, L=2, \alpha=\frac{0.3}{L^2}$ are shown.
The hierarchy between the bunch of spectra near the origin and the others is clear
for the lowest three momentum sectors. By counting the numbers of the spectra 
surrounded by the circles in the figure, 
the excess over the formula \eq{eq:numzero} is obtained as 
10 at $p=0$ and 6 at $p\neq 0$.

\section{A physical interpretation of the very light modes}
\label{sec:interpretation}
In the preceding two sections, 
it is observed that the distribution of the very light modes 
has a characteristic feature depending on the dimensions, which is summarized
in Table~\ref{tab:dist}. 
\begin{table}
\begin{center}
\begin{tabular}{|c|c|c|}
\hline
$D$ & {\#}light modes at $p=0$ & {\#}light modes at $p\neq 0$ \\
\hline
1&1&0 \\
\hline
2&3&1\\
\hline
3&6&3 \\
\hline
4&10&6 \\
\hline
\end{tabular}
\end{center}
\caption{The distribution of the very light modes obtained from the numerical analysis. }
\label{tab:dist}
\end{table} 
A general expectation in lattice-like approaches is that,
in the continuum limit $L\rightarrow\infty$, the low-energy effective
physics is governed by such light modes, while the ``heavy" spectra 
$\gtrsim O(0.1)$ appear only at a ``Planck" energy, and
the zero modes are just the unphysical gauge modes. 
In this section, it will be pointed out 
that the distribution of the very light modes can easily be explained by
the general relativity.

The fundamental variable of the general relativity is a metric tensor. Since 
the general relativity has the gauge symmetry of the general coordinate transformation,
the real degrees of freedom are the metric tensor modulo
the general coordinate transformation. 
These real degrees of freedom should be compared with the 
numerical result. As mentioned previously, the background space corresponding 
to the solution \eq{eq:cpgp} with \eq{eq:a} is a fuzzy $D$-dimensional flat torus. 
The infinitesimally small general coordinate transformation around
a flat space background is given by
\be
\delta g_{\mu\nu}=\partial_\mu v_\nu+ \partial_\nu v_\mu,
\ee  
where $v_\mu$ is the local translational transformation vector. 
In the momentum basis, this is
\be
\delta g_{\mu\nu}(p)=i p_\mu v_\nu(p) +i p_\nu v_\mu(p).
\ee
Therefore, in the $p=0$ sector, the gauge transformation does not generate any 
transformations, while, in the $p\neq 0$ sectors, it generates a $D$ dimensional 
transformation.
Thus the degrees of the freedom of the metric tensor modulo the general 
coordinate transformation are given by 
\be
\# g_{\mu\nu}/{\rm g.c.t.}=\left\{ 
\begin{array}{cl}
\frac{D(D+1)}{2}& {\rm for}\ p=0, \\
\frac{D(D-1)}{2} & {\rm for} \ p\neq 0. 
\end{array} 
\right.
\ee
This formula exactly agrees with Table~\ref{tab:dist}.

\section{Discussions and future prospects}
In this paper, numerical analysis of the fluctuation spectra around a Gaussian 
classical solution of a tensor model was carried out. 
It was observed that the distribution of the very light modes, 
which are expected to be the low-energy effective physical modes 
in the continuum limit,  
agrees exactly with what is expected from the general relativity in one to four dimensions.
This result suggests that the low-energy effective field theory of the tensor model 
around the classical solution can be described in a similar manner as
the general relativity.

The exact agreement in various dimensions would be a good evidence of 
the link between the tensor model and the general relativity, but
it is clear that more evidences are needed to definitely prove it.
A direction would be to study the properties of the very light modes, and compare 
them with the general relativity.
For this purpose, the very light modes must be distinguished from the 
other modes, especially from the zero modes. 
This can be done with the correct numerical method employed in Section~\ref{sec:flcnum}, 
but not with the efficient but approximate method in Section~\ref{sec:distribution},
since, in the latter method,
the very light modes and the zero modes form a bunch. 
Therefore, with the present computational facility
mentioned in Section~\ref{sec:flcnum},  
the properties of the very light modes 
can be studied only in one and two dimensions for small $L$. 

To study the properties of the very light modes in larger $D,L$,
it would be necessary to develop a systematic method to reduce the 
number of the degrees of freedom of the tensor model without affecting the 
properties of the very light modes. This would also be necessary to study 
the quantum properties of the tensor model.
One candidate for such a method would be a kind of renormalization procedure
for the tensor model. 
A key issue would be the treatment of the zero modes, which can
easily mix with the very light modes in any rough approximate method 
like that in Section~\ref{sec:distribution}.
It would be important to keep the symmetry of the tensor model, 
so that the zero modes be 
distinguishable from the very light modes.
 
Besides the huge number of the degrees of freedom of the dynamical variables, 
what makes tensor models quite complicated is
the lack of a guiding principle to chose among the infinite possibilities
of actions.
It would be necessary to single out the elements essential 
to induce the general relativity at low energy, and 
get a hind for a principle.
It would be also important to investigate the universality classes of tensor models,
to see whether the result of this paper is just a special case or universal in tensor
models.

Finally
it would be amusing to see that some constructive approaches
to string/M theory \cite{Banks:1996vh,Ishibashi:1996xs} have dynamical variables ${(M_\mu)^a}_b$, which 
is similar to a rank-three tensor. 
A tensor model is discussed also as a toy model of open membrane theory \cite{Ho:2007vk}.
Thus, as explained in Section~\ref{sec:intro},
tensor models appear in various approaches to quantum gravity, i.e. 
dynamical triangulation, loop quantum gravity, fuzzy spaces and string/M theory. 
It would be highly interesting, if tensor models connect
all of these apparently different approaches in future.  
 
\vspace{.5cm}
\section*{Acknowledgments}
The author would like to thank H.~Kawai for some useful comments on the occasion of 
a seminar at the preliminary stage of this work.
The author was supported in part by the Grant-in-Aid for Scientific Research No.13135213, No.16540244 and No.18340061
from the Ministry of Education, Science, Sports and Culture of Japan.

\appendix

\section{The computation of the partial derivatives }
\label{ap:A}
In this appendix,  the explicit formula of the partial derivatives in \eq{eq:dsdcdc} 
are given.

The partial derivative of the first term of $W_{abc}$ in \eq{eq:defofw} is obvious,
\be
\frac{\partial C_{p_4p_5p_6}}{\partial C_{p_1p_2p_3}}=\delta_{p_4p_1} \delta_{p_5p_2} \delta_{p_6p_3}. 
\ee
Here the permutation symmetry of the indices of the tensor $C_{abc}$ 
will be cared by the summation over the permutations $\sigma,\sigma'$ in \eq{eq:dsdcdc}.
The same kind of simplification of expressions will also be used in the formulas below.

\begin{figure}
\begin{center}
\includegraphics[scale=0.7]{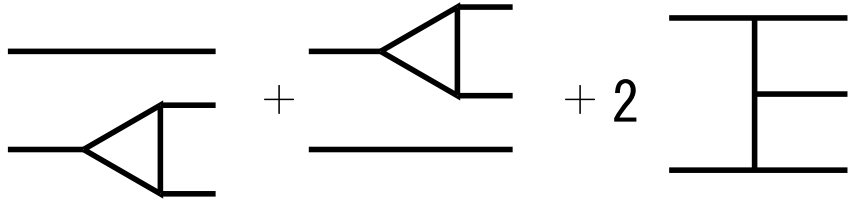}
\end{center}
\caption{The partial derivative $\frac{\partial K_{ab}}{\partial C_{cde}}$. 
The open ends in the
left are associated with the indices of $K_{ab}$, and those of $C_{cde}$
are in the right. 
The permutation symmetry of the indices of $C_{abc}$ will be cared by the summation of 
$\sigma,\sigma'$ in \eq{eq:dsdcdc}.}
\label{fig:km1}
\end{figure}

The partial derivative of the second term of $W_{abc}$ consists of two parts.
One is the derivative
$\frac{\partial K^{-1}_{ab}}{\partial C_{cde}}$.
This can be done through the well known formula,
\be
\frac{\partial K^{-1}_{ab}}{\partial C_{cde}}=-K^{-1}_{aa'}
\frac{\partial K^{a'b'}}{\partial C_{cde}}
K^{-1}_{b'b}.
\ee 
The definition of $K_{ab}$ is given by neglecting the bar $\bar\ $ 
in \eq{eq:kp}. From this definition,  
the derivative $\frac{\partial K^{ab}}{\partial C_{cde}}$
can be obtained as the diagram in Figure~\ref{fig:km1}.   

\begin{figure}
\begin{center}
\includegraphics[scale=0.7]{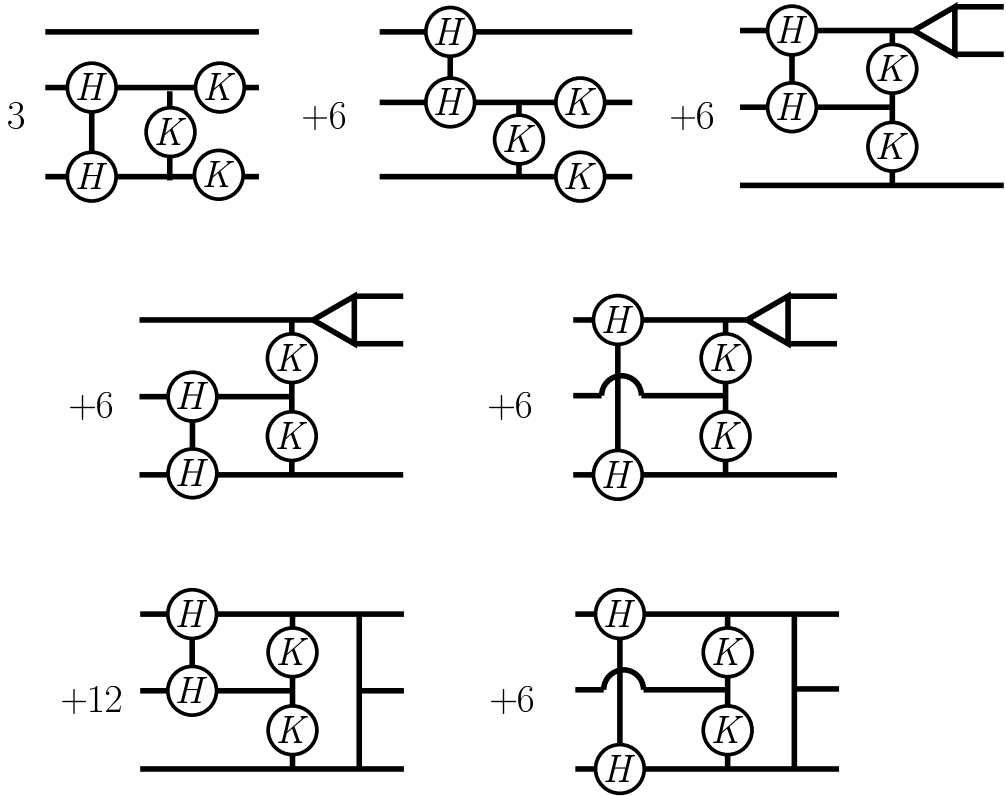}
\end{center}
\caption{The partial derivative $\frac{\partial I_{abc}}{\partial C_{def}}$. 
The open ends in the
left are associated with the indices of $I_{abc}$, and those of $C_{def}$ are 
in the right.
The permutation symmetry of the indices of $I_{abc}$ 
must be cared by averaging over all the permutations of the open ends in the left,
while the permutation symmetry of the indices of $C_{abc}$ are 
cared by the summation of $\sigma,\sigma'$ in \eq{eq:dsdcdc}.}
\label{fig:idc}
\end{figure}

The other part is the partial derivative $\frac{\partial I_{abc}}{\partial C_{def}}$. 
The definition of $I_{abc}$ is given by neglecting the bar $\bar\ $ in \eq{eq:ip}. 
The result of the partial derivative is shown in Figure~\ref{fig:idc}. 
In this figure, to avoid a large figure, 
it is omitted to average over all the permutations of the open ends 
in the left, which is needed to take into account 
the permutation symmetry of the indices of $I_{abc}$. 

\section{The raw numerical results}
\label{ap:B}

\subsection{ The numerical results for $D=1$, $L=3$}
Because of the permutation symmetries of the indices of $C_{p_1p_2p_3}$ 
and the ansatz, there are only 6 independent
variables in the optimization process. 
The minimum of \eq{eq:fc} reached by the Nelder-Mead method 
is $f(C) \sim 2\times 10^{-25}$. The values of the six $C^0_{p_1p_2p_3}$ 
at the minimum are listed in Table~\ref{tab:L3}. 
\begin{table}
\begin{center}
\begin{tabular}{|c|c|c|}
\hline
-3 0 3 & -3 1 2 & -2 0 2 \\
\hline 
1.84462617663699010E-001& 2.40366883259875430E-001& 3.61070116973867590E-001\\
\hline \hline
 -2 1 1& -1 0 1& 0 0 0 \\
 \hline
4.17490143455173490E-001& 5.58142487944932600E-001& 6.53570115178405460E-001 \\
\hline
\end{tabular}
\end{center}
\caption{The solution $C^0_{p_1p_2p_3}$ for $D=1,L=3$. 
The momenta $p_1p_2p_3$ are listed in the upper colums, 
and the corresponding values of $C^0_{p_1p_2p_3}$ 
are in the lower.}
\label{tab:L3}
\end{table}

In Table~\ref{tab:L3spec}, the fluctuation spectra obtained by diagonalizing \eq{eq:dsdcdc} are 
shown for each momentum sector.
\begin{table}
\begin{center}
\begin{tabular}{|c|c|c|}
\hline 
$p$ & spectra & \#{zero} \\
\hline
& 
-2.074144E-015 1.296988E-015 2.010206E-003 1.187204E+000  
& \\
0&
2.226274E+000 6.450744E+000 2.101532E+001 1.343863E+003&
 2\\
\hline
&
-8.779625E-014 -1.021941E-014 -1.835845E-015 1.208359E+000& \\
1&1.623874E+000 3.193161E+000 2.674381E+001 1.130220E+003&
3 \\
\hline
 & 
-5.557866E-015 5.578683E-015 1.018426E+000 1.622450E+000& \\
2&3.177940E+000 3.086634E+001 6.450256E+002 &
2\\
\hline
 &
-1.110795E-014 -3.234454E-016 9.040864E-001 1.751722E+000& \\
3&2.161799E+000 2.207987E+001 2.123835E+002 &
2\\
\hline
&
-5.834977E-015 5.523345E-001 1.344606E+000 4.763487E+000 & \\
4&4.064315E+001&
1\\
\hline
5 &
1.792788E-016 3.732453E-001 1.315828E+000 1.723215E+001 
& 1 \\
\hline
6 &
6.317254E-001 9.912976E-001 7.289011E+000 
& 0 \\
\hline
7 &
4.695151E-001 1.014654E+000 
& 0\\
\hline 
8 &
9.269276E-001 
& 0\\
\hline
9 &
9.965088E-001 
& 0\\
\hline
\end{tabular}
\end{center}
\caption{The values of the fluctuation spectra obtained by diagonalizing \eq{eq:dsdcdc} for $D=1$, $L=3$.
The momentum of the modes, the fluctuation spectra, 
and the number of the zero modes from the formula \eq{eq:numzero}
are listed for each momentum sector.
The negative $p$ sectors are omitted as obvious, since the ansatz assumes the reflection symmetry 
$p\rightarrow -p$.}
\label{tab:L3spec}
\end{table} 
The spectra contains very tiny vales $\lesssim 10^{-14}$. 
These can be identified with the zero modes from the symmetry 
breaking $SO(n,R)/SO(2)$, since the numbers agree with the formula \eq{eq:numzero}.
This agreement is also a good check for the correctness of the computational 
codes, because any slight mistakes in the codes will break this agreement.

There exists a non-zero but a very small spectrum $ \sim 2\times10^{-3}$
only at $p=0$ sector.
The other spectra are $\gtrsim O(0.1)$.

\subsection{The numerical results for $D=1$, $L=5$}
The minimum value reached is $f(C)\sim 3\times 10^{-25}$. The solution is shown in Table~\ref{tab:L5sol}.
\begin{table}
\begin{center}
\begin{tabular}{|c|c|c|}
\hline
-5 0 5 & -5 1 4 &-5 2 3 \\
\hline 
5.30666376133652470E-002 & 7.72763400282020090E-002& 9.34623815478460020E-002 \\ \hline\hline
-4 0 4 & -4 1 3 &-4 2 2 \\
\hline
1.24046348049658310E-001& 1.65453909450288150E-001& 1.82235408294834920E-001 \\ \hline\hline
-3 0 3 &-3 1 2 &-2 0 2 \\
\hline
2.43404073757127580E-001 & 2.95609691797031670E-001& 3.95946657504184290E-001 \\ \hline\hline
 -2 1 1& -1 0 1& 0 0 0 \\
 \hline
 4.36599185858296160E-001& 5.30865955626977360E-001& 5.85474131045624180E-001\\
\hline
\end{tabular}
\end{center}
\caption{The table of the solution $C^0_{p_1p_2p_3}$ for $D=1$, $L=5$. }
\label{tab:L5sol}
\end{table}
The fluctuation spectra at the $p=0$ sector are shown in Table~\ref{tab:L5spec}.
\begin{table}
\begin{center}
\begin{tabular}{|c|c|c|}
\hline
$p$ & spectra & \#{zero} \\
\hline
& 
-2.796163E-015 -8.151607E-016 2.246353E-016 1.433114E-015 &\\
&
1.073950E-006 9.152090E-001 9.941203E-001 1.015528E+000 &\\
0 &
1.037451E+000 1.116800E+000 1.242951E+000 1.383553E+000 &
4 \\
&
1.934481E+000 4.332078E+000 5.159299E+000 8.182407E+000 &\\
&
2.407250E+001 1.363542E+003 
&\\
\hline
\end{tabular}
\end{center}
\caption{The fluctuation spectra in the $p=0$ sector for $D=1$, $L=5$, and the 
number of the zero modes from \eq{eq:numzero}. }
\label{tab:L5spec}
\end{table}
There the tiny spectra $\lesssim 10^{-15}$ can be identified with the zero modes, the number of which
agrees with the formula \eq{eq:numzero}.
There exists a very light mode $\sim 10^{-6}$, and the others are $\gtrsim O(1)$.
The fluctuation spectra at the other momentum sectors are also obtained, but the values are 
not listed here. The spectra are plotted in Figure~\ref{fig:Lspec} instead. 
At the $p\neq 0$ sectors,
the fluctuation spectra can be divided into two parts. One part 
consists of the tiny spectra $\lesssim 10^{-13}$, 
the numbers of which agree with \eq{eq:numzero}. The other consists of the ``
heavy" modes with the spectra $\gtrsim O(0.1)$.
  
\subsection{The numerical results for $D=1$, $L=10$}
The number of independent variables is 36, and the minimizing process takes a longer time.
A one-day run was carried out, but a minimum vanishing within the 
computational accuracy was not reached. 
The minimizing process was terminated at $f(C)\sim  10^{-14}$.   
The fluctuation spectra are plotted in Figure~\ref{fig:Lspec}.
The qualitative features are the same as $L=3,5$. 
A clear agreement between the number of the tiny spectra and 
\eq{eq:dsdcdc} is obtained. 
There exists again a very light mode only at the $p=0$ sector, which
has a spectrum $\sim 3 \times 10^{-7}$.       
      
\subsection{The numerical results for $D=2$, $L=2$}
The minimum value reached is $f(C)\sim 10^{-25}$. The values of $C_{abc}^0$
at the minimum are shown in Table~\ref{tab:cd2l2}.
\begin{table}
\begin{center}
\begin{tabular}{|c|c|c|}
\hline
(-2,-2) (0,0) (2,2)&(-2,-2) (0,1) (2,1)&(-2,-2) (0,2) (2,0)\\ 
\hline
1.33090366933366510E-001&
1.54486879640217570E-001&
1.33090366933252490E-001\\
\hline
\hline
(-2,-2) (1,1) (1,1)&(-2,-1) (0,0) (2,1)&(-2,-1) (0,1) (2,0)\\
\hline
1.79323241274937400E-001&
2.04924589321128730E-001&
2.04924589320929830E-001\\
\hline
\hline
(-2,-1) (0,2) (2,-1)&(-2,-1) (1,-1) (1,2)&(-2,-1) (1,0) (1,1)\\
\hline 
1.54486879639913640E-001&
1.79323241274623960E-001&
2.37869660255753430E-001\\
\hline
\hline
(-2,0) (0,0) (2,0)&(-2,0) (1,-1) (1,1)&(-2,0) (1,0) (1,0)\\
\hline
2.54117050163986060E-001&
2.37869660255717370E-001&
2.94970635725226470E-001\\
\hline
\hline
(-1,-1) (0,0) (1,1)&(-1,-1) (0,1) (1,0)&(-1,0) (0,0) (1,0)\\
\hline
3.15530629872480820E-001&
3.15530629872307290E-001&
3.91274239784655140E-001\\
\hline
\hline
(0,0) (0,0) (0,0)&&\\
\hline
4.85200219013368070E-001&
&
\\
\hline
\hline 
\end{tabular}
\end{center}
\caption{The solution $C^0_{(p_1^1,p_1^2)(p_2^1,p_2^2)(p_3^1,p_3^2)}$ for $D=2$, $L=2$.}
\label{tab:cd2l2} 
\end{table}
The fluctuation spectra at all the momentum sectors 
are computed by diagonalizing \eq{eq:dsdcdc}. 
To avoid a huge table, 
only the lower parts of the spectra in the $p=(0,0)$ and $p=(1,0)$ sectors 
are shown in Table~\ref{tab:specd2l2}. 
There the number of modes with values $\lesssim 10^{-14}$
agree with the formula \eq{eq:numzero}, and can be identified with the zero modes. 
There seem to exist a number of very light modes. In the $p=(0,0)$ sector, 
the three modes with 
$\sim 8\times 10^{-9},3\times 10^{-3},3\times 10^{-2}$ may be identified with the 
very light modes.
In the $p=(1,0)$ sector, the mode with $\sim 4\times 10^{-2}$ may be identified with 
a very light mode. 
The identification of the very light modes is not so clear in this case, but will be 
improved much more in $L=3$.
The spectra in all the other momentum sectors are computed as well, but are not shown here.
There the modes with spectra $\lesssim 10^{-13}$ can clearly be identified
with the zero modes, and their numbers agree with the formula \eq{eq:numzero}. There the very light modes and the modes with spectra $\gtrsim O(0.1)$ cannot be well distinguished.
\begin{table}
\begin{center}
\begin{tabular}{|c|c|c|}
\hline
$p$
&
spectra
&
\#{zero} 
\\
\hline
&
-3.422193E-015 -2.176675E-015 -8.331255E-016 -2.168759E-016 
&
\\
&
5.312991E-017 1.245100E-016 6.100288E-016 1.548574E-015 
&
\\
(0,0)&
3.822409E-015 6.586865E-015 8.393766E-009 2.734072E-003 
&
10\\
&
2.650856E-002 1.375314E-001 5.215319E-001 5.803728E-001 
&
\\
&
5.930927E-001 7.295304E-001 $\cdots$ &
\\
\hline
&
-2.247079E-014 -6.834805E-015 -3.502970E-015 -1.232192E-015 
&
\\
&
-6.677623E-016 6.984656E-017 1.852255E-016 3.004076E-015 
&
\\
(1,0)&
5.622760E-015 1.195054E-014 4.019257E-002 2.451403E-001 
&
10
\\
&
2.749515E-001 5.590763E-001 5.978878E-001 $\cdots$
&
\\
\hline
\end{tabular}
\end{center}
\caption{The lower part of the fluctuation spectra at the $p=(0,0),(1,0)$ sectors for $D=2$, $L=2$, and the number of the zero modes \eq{eq:numzero}. } 
\label{tab:specd2l2}
\end{table} 

\subsection{The numerical results for $D=2$, $L=3$}
The number of independent variables is 42. The minimum value reached is $f(C)\sim 10^{-22}$.
To avoid a large table, the solution is not explicitly shown in this case. 
The fluctuation spectra are computed in all the momentum sectors. 
Some parts of the spectra in some low momentum sectors are shown 
in Table~\ref{tab:specd2l3}.
In the $p=(0,0)$ sector,
the formula \eq{eq:numzero} implies that the modes with spectra $\lesssim 10^{-14}$ 
should be regarded as the zero modes. Then the very light modes 
can be identified with
those with the spectra $1.3\times 10^{-12}$, $3.0\times 10^{-4}$ and $1.1\times 10^{-3}$.
In the non-vanishing momentum sectors, the number of the modes with spectra 
$\lesssim 10^{-13}$ agrees with the formula \eq{eq:numzero}.  

\begin{table}
\begin{center}
\begin{tabular}{|c|c|}
\hline
&
-7.023228E-015 -3.339147E-015 -1.354682E-015 -1.152946E-015
\\
&
-1.022040E-015 -8.653429E-016 -5.793971E-016 -4.444370E-016
\\ 
&
-2.825529E-016 -1.526719E-016 -9.835320E-017 7.877653E-017
\\
(0,0)& 
2.207186E-016 3.057724E-016 4.016902E-016 5.873527E-016
\\
&
8.141446E-016 1.301374E-015 1.455441E-015 2.779198E-015
\\
&
5.143835E-015 1.357294E-014 1.312193E-012 3.048268E-004
\\
&
1.069215E-003 1.385427E-001 3.174581E-001 $\cdots$
\\
\hline
(1,0)
&
$\cdots$ 3.531105E-014 2.117237E-003 2.150369E-001 $\cdots$
\\
\hline
(1,1)
&
$\cdots$ 2.769432E-014 9.569329E-004 1.232058E-001 $\cdots$
\\
\hline
(2,0)
&
$\cdots$ 6.895682E-015 6.938484E-003 2.403594E-001 $\cdots$
\\
\hline
\end{tabular}
\end{center}
\caption{A part of the fluctuation spectra for $D=2$, $L=3$. For the $p=(0,0)$ sector, all the lower spectra are shown. For $p=(1,0),(1,1),(2,0)$ sectors, 
only the very light modes and the nearby spectra are shown.}
\label{tab:specd2l3}
\end{table}

\section{Analytical expressions of the loop integrals in $\left. \frac{\partial W_{abc}}{\partial C_{def}}\right|_{C=\bar C}$}
\label{app:gauss}
\begin{figure}
\begin{center}
\includegraphics[scale=.7]{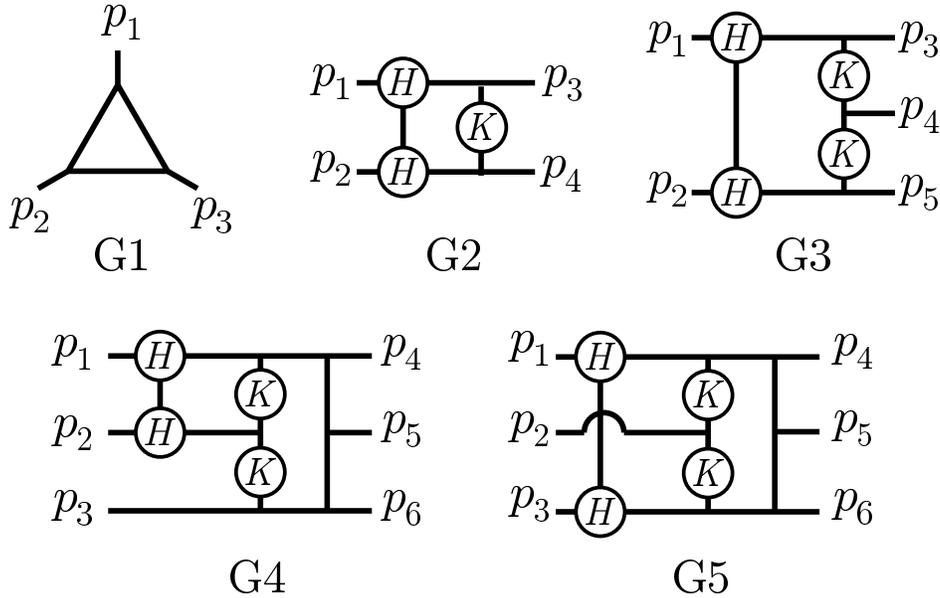}
\end{center}
\caption{The ingredients for evaluating $\left. \frac{\partial W_{abc}}{\partial C_{def}}\right|_{C=\bar C}$. The momenta flow inward. }
\label{fig:ing}
\end{figure}

In this appendix, the analytical results of the loop integrals necessary 
for evaluating $\left. \frac{\partial W_{abc}}{\partial C_{abc}}\right|_{C=\bar C}$ are given.
The diagrams in Figure~\ref{fig:km1} and \ref{fig:idc} show that the kinds of
loop integrals can be summarized as in Figure~\ref{fig:ing}.

The explicit analytical expressions for these diagrams are given in the followings.
The expressions must be multiplied by the obvious delta functions of momentum conservation, which are omitted below for simplification, 
e.g. $\delta^D(p_1+p_2+p_3)$ for G1, etc. 
\bea
{\rm G1:}&&A^3
\left(\frac{\pi}{6\alpha}\right)^{\frac{D}{2}} 
\exp\left[-\frac{5\alpha}{3}\left(p_1^2 + p_2^2 + p_3^2\right)\right].
\\
{\rm G2:}&&
A^{36}
\left(\frac{\pi}{4 \sqrt{2} \alpha}\right)^{7D} 
\left(\frac{\pi}{18\alpha}\right)^D \left(\frac{\pi}{20\alpha}\right)^{D/2} 
\exp\left[ -\alpha \left(3\left(p_1^2 +
       p_2^2\right) + 5 p_3^2 + 5 p_4^2  \right. \right.\cr
       &&
\left.\left.
-\frac{1}{20}\left(-3 p_1 + 3p_2 - 
      7p_3 + 7p_4\right)^2  + 3\left(p_2 + p_4\right)^2 
+ 3\left(p_1 + p_3\right)^2\right)\right].
\\
{\rm G3:}&&
A^{41}
\left(\frac{\pi}{4 \sqrt{2}\alpha}\right)^{8D} \left(\frac{\pi}{18\alpha}\right)^D
\left(\frac{\pi}{26 \alpha}\right)^{\frac{D}2} 
\exp \left[
-\alpha \left(
3 p_1^2 + 3 p_2^2 + 5 p_3^2 + 7
      p_4^2 + p_5^2 \right.\right. \cr
      &&
\left.\left. - \frac{2}{13}\left(-3 p_1 - 
    5 p_3 + 5 p_4 + 2p_5\right)^2+ 4 (p_4 + p_5)^2 + 6 (p_1 + p_3)^2
\right)\right].\\
{\rm G4:}&&
A^{44}
\left(\frac{\pi}{4\sqrt{2}\alpha}\right)^{8D}\left(\frac{\pi}{18\alpha}\right)^D\left(\frac{\pi^2}{4\cdot 7\cdot 13 \alpha^2}\right)^{\frac{D}{2}}
\exp\left[-\frac{\alpha}{91}\left( 749 p_1^2 + 821p_2^2 + 421p_3^2 + 50p_3p_4 \right.\right.\cr
&&
+ 177p_4^2 + 
     132p_3p_5 + 10p_4p_5 + 177p_5^2 + 396p_3p_6 + 30p_4p_6 + 152p_5p_6 + 
     319p_6^2 + \cr
&&
\left.\left.
 14p_1(47p_2 + 15p_3 + 10p_4 + 3p_5 + 9p_6) + 
     2p_2(150p_3 + 61p_4 + 30p_5 + 90p_6)\right)\right].
     \\
{\rm G5:} &&A^{44}
\left(\frac{\pi}{4\sqrt{2}\alpha}\right)^{8D}\left(\frac{\pi^2}{18\alpha^2}\right)^{D}\left(\frac{1}{8\cdot 47}\right)^\frac{D}{2}
\exp\left[-\frac{\alpha}{188}\left(1499 p_1^2 + 1011 p_2^2 + 1283 p_3^2 + 370p_3p_4 
\right. \right.
\cr
&&
+ 803p_4^2 + 222p_3p_5 + 738p_4p_5 + 
635p_5^2 + 74p_3p_6 + 246p_4p_6 + 298p_5p_6 + 363p_6^2 + \cr
&&
\left. \left.
10p_2\left( 41p_3 + 11(5p_4 + 3p_5 + p_6)\right) + 
     2p_1\left( 313p_2 + 443p_3 + 73(5p_4 + 3p_5 + p_6)\right) \right) \right].
\eea

\end{document}